\documentclass[twocolumn,showpacs,preprintnumbers,amsmath,amssymb]{revtex4}

\usepackage{graphicx}% Include figure files
\usepackage{dcolumn}% Align table columns on decimal point
\usepackage{bm}% bold math
\usepackage{graphicx}
\usepackage{epsfig}
\include{graphic}

\begin{document}
\preprint{APS/123-QED}
\title{Quantized Dispersion of Two-Dimensional Magnetoplasmons Detected by Photoconductivity Spectroscopy}

\author{S. Holland$^{1}$, Ch. Heyn$^{1}$, D. Heitmann$^{1}$, E. Batke$^{2}$, R. Hey$^{3}$, K. J. Friedland$^{3}$, and C. -M. Hu$^{1,}$\footnote{Electronic address: hu@physnet.uni-hamburg.de}}

\affiliation{$^{1}$Institut f\"ur Angewandte Physik und Zentrum
f\"ur Mikrostrukturforschung, Universit\"at Hamburg,
Jungiusstra\ss e 11, 20355 Hamburg, Germany}
\affiliation{$^{2}$Physikalisches Institut der Universit\"at
W\"urzburg, Am Hubland, D-97074 W\"urzburg, Germany}
\affiliation{$^{3}$Paul-Drude-Institut f\"ur
Festk\"orperelektronik, Hausvogteiplatz 5-7, D-10117 Berlin,
Germany}
\date{\today}

\begin{abstract}

We find that the long-wavelength magnetoplasmon, resistively
detected by photoconductivity spectroscopy in high-mobility
two-dimensional electron systems, deviates from its well-known
semiclassical nature as uncovered in conventional absorption
experiments. A clear filling-factor dependent plateau-type
dispersion is observed that reveals a so far unknown relation
between the magnetoplasmon and the quantum Hall effect.

\end{abstract}


%72.25.Fe Optical creation of spin polarized carriers
%72.25.Rb Spin relaxation and scattering
%73.20.Mf
%73.43.-f Quantum Hall effects
%73.43.Lp Collective excitations
%73.43.Qt Magnetoresistance
%73.50.Jt Galvanomagnetic and other magnetotransport effects (including thermomagnetic effects)
%73.50.Mx High-frequency effects; plasma effects
%73.50.Pz Photoconduction and photovoltaic effects
%73.63.Kv Quantum dots
%78.30.-j Infrared and Raman spectra
%78.67.Hc Quantum dots

\maketitle Photoconductivity experiments on two-dimensional
electron system (2DES) have recently attracted enormous interest
triggered off by the discovery of the zero resistance states in
extremely high-mobility samples in the microwave regime
\cite{Mani}. Despite of the latest progress of resistively
detection of the edge magnetoplasmon \cite{Kukushkin}, the role of
the magnetoplasmon has been pointed out to be troublesome
\cite{Mikhailov}. The question is whether the charge excitations
detected resistively are identical to what we know from the
absorption experiments, as one might naively anticipate. Indeed,
carefully reanalyzing the original data in Ref. 1 revealed
surprisingly a shifted cyclotron resonance (CR) frequency
$\omega_{c}=eB/m^{\ast}$ with a reduced effective mass $m^{\ast}$
\cite{Zudov}. The above critical question is not clarified because
directly comparing photoconductivity and absorption spectroscopy
in the microwave regime is a cumbersome task.

At first glance, two-dimensional plasmon seems to be unlikely a
subject to give us surprises. Its dispersion in the
long-wavelength limit was predicted as early as in 1967 by Stern
\cite{Stern} with
\begin{equation}
\omega^{2}_{p}(q)=\frac{N_{s}e^{2}}{2\varepsilon
\varepsilon_{0}m^{\ast}}q,
\end{equation}
which describes the collective charge oscillation of a 2DES with
charge density $N_{s}$ and effective mass $m^{\ast}$ at the wave
vector $q$ oriented in the 2D plane. The effective dielectric
permittivity $\varepsilon(q)$ depends on the surrounding medium.
In the presence of a perpendicular magnetic field the
magnetoplasmon frequency $\omega_{mp}$ is given by \cite{Chiu}
\begin{equation}
\omega^{2}_{mp}(B)=\omega^{2}_{p}(q)+\omega^{2}_{c}.
\end{equation}
Both Eqs. (1) and (2) have been confirmed by many experiments
\cite{Batke}, which makes the plasmon a very well understood
elementary excitations of the 2DES.

By combining eq. (1) and (2) it is straightforward to define a
renormalized magnetoplasmon frequency $\Omega_{mp}$ and find
\begin{equation}
\Omega_{mp}\equiv\frac{\omega^{2}_{mp}(B)-\omega^{2}_{c}}{\omega_{c}}=\frac{\hbar\cdot
q_{TF}\cdot q}{2m^{\ast}}\nu,
\end{equation}
where we define $q_{TF} = m^{\ast}e^{2}/2\pi \varepsilon
\varepsilon_{0}\hbar^{2}$ as the effective Thomas-Fermi wave
vector depending on $\varepsilon(q)$. The monotonous linear
dependence of $\Omega_{mp}$ on the filling factor $\nu=hN_{s}/eB$
emphasizes the semiclassical nature of the magnetoplasmon, because
Eq. (2) was obtained by analyzing the self-consistent response of
the 2DES to a longitudinal electric field in the semiclassical
limit, in which the quantum oscillatory part of the polarizability
tensor was disregarded \cite{Chiu}. The main result we will
present in this paper is an astonishing deviation of $\Omega_{mp}$
from Eq. (3) for resistively detected magnetoplasmons in
high-mobility 2DESs, where we find a quantized dispersion with
plateaus forming around even filling factors. It reveals a
previously unknown relation between the magnetoplasmon and the
integer quantum Hall effect (QHE) which is intriguing for
investigating the nature of both.

%\begin{figure}
%\begin{center}
%%\hskip -0.7cm
%\epsfig{file=Fig1.eps,height=3.5cm} \caption{Schematic view on the
%sample structure and bias circuit showing a meandering long Hall
%bar with ohmic contacts and a grating coupler.} \label{fig_Fig1}
%\end{center}
%\end{figure}

To explore a wide range of filling factors we choose a high
mobility 2DES with high density confined in a GaAs quantum well
\cite{Friedland}, \textit{cf.} the sample M1218 in Table 1, and we
compare it with 2DESs with different mobilities either formed at
the interface (HH1295) or in a quantum well (M1266). Extremely
long 2DES Hall bars with the length $L$ of about 0.1 m and the
channel width $W$ of about 40 $\mu$m were defined by chemical wet
etching. Ohmic contacts were made by depositing AuGe alloy
followed by annealing. A gold grating coupler with a period of $a$
= 1 $\mu$m was fabricated on top of the meandering Hall bar
perpendicular to the current path, which allows us to couple the
2D plasmon at the wave vectors $q = 2\pi\cdot n/a$ ($n$ = 1, 2,
...) with THz radiation. In this frequency regime, the excitations
measured by far-infrared photoconductivity (FIR-PC) and absorption
spectroscopy can be directly compared. All data presented in this
paper were measured at 1.8 K in the Faraday geometry. Details of
our experimental set up have been published elsewhere \cite{Hu}.

\begin{table}[h]
\begin{center}
\label{munstest} \caption{Parameters of the samples. The two
$q_{TF}$ values for sample HH1295 are obtained for the $n$=1
($n$=2) plasmon mode, respectively.}
\begin{tabular}{ccccc}
  \hline\hline    Sample & $\mu$ & $N_{s}$ & $ m^{\ast}$ & $ q_{TF}$ \\
         &(10$^{6}$cm$^{2}$/Vs) & (10$^{11}$cm$^{-2}$) & ($m_{e}$) & (10$^{6}$cm$^{-1}$) \\
  \hline    M1218 & 1.3 & 5.58 & 0.0726 & 1.83 \\
            HH1295 & 0.5 & 1.93 & 0.0695 & 1.55 (1.86) \\
            M1266 & 0.3 & 7.18 & 0.0730 & 1.94 \\
  \hline\hline
\end{tabular}
\end{center}
\end{table}

\begin{figure}
\begin{center}
\epsfig{file=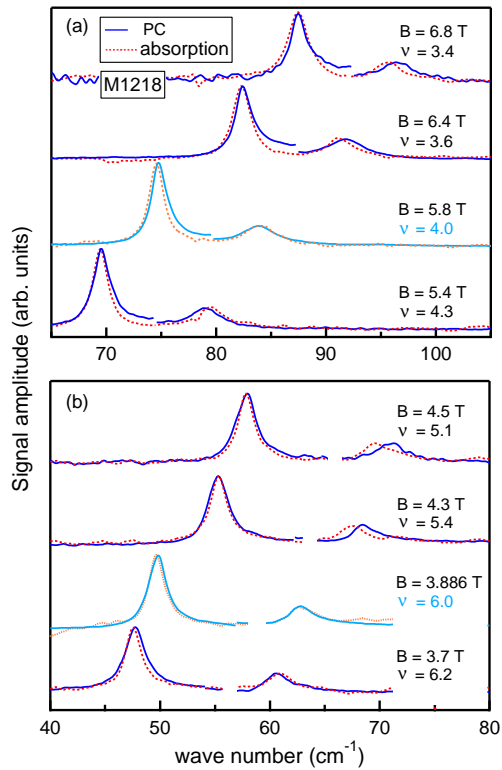,height=10.5cm} \caption{(color online).
FIR-PC spectra (solid curves) measured on sample M1218 around (a)
$\nu$ = 4 and (b) $\nu$ = 6 in comparison with absorption spectra
(dotted curves), displaying $\nu$-dependent variations of the
magnetoplasmon energy detected in the PC spectra.} \label{Fig2}
\end{center}
\end{figure}

In Fig. \ref{Fig2} we plot and compare FIR-PC (solid curves) and
absorption spectra (dotted curves) measured on sample M1218 around
$\nu$ = 4 and 6. The dominant resonance at the lower energy is the
CR, while the weak resonance at the higher energy side is the
magnetoplasmon at $q = 2\pi/a$ = 6.28$\times 10^{4}$ cm$^{-1}$. In
contrast to the absorption spectroscopy which probes the
high-frequency conductivity of the 2DES so that the resonant
strength is determined by the transition matrix element and the
electron density, the sensitivity for FIR-PC depends strongly on
the filling factor \cite{Hu}. For better comparison of the
excitation energy, which is the focus of this paper, we normalize
the CR and magnetoplasmon in the FIR-PC spectra in Fig. \ref{Fig2}
so that they are displayed at the same level of that in the
absorption spectra. At the even integer filling factors $\nu$ = 4
and 6, no deviation of the magnetoplasmon energy is found. By
increasing (decreasing) the $B$ field, the resistively detected
magnetoplasmon shifts to higher (lower) energy compared to that in
the absorption spectra. No such changes in the CR energy are
found, except that the CR line shape shows slight deviations in
the higher-energy side.

\begin{figure}
\begin{center}
\hskip -0.7cm\epsfig{file=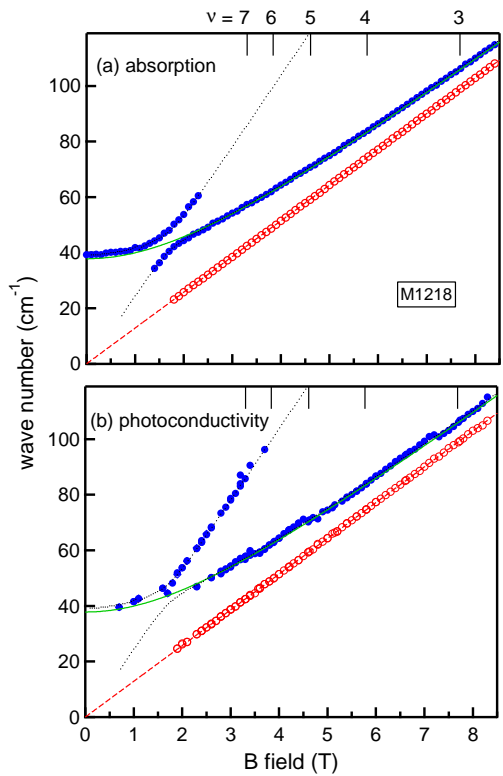,height=10.5cm} \caption{(color
online). $B$-field dispersion of the CR and magnetoplasmons
measured in sample M1218 by (a) absorption and (b) FIR-PC
spectroscopy. In (a) the CR frequency is fitted by the relation
$\omega_{c} = eB/m^{\ast}$ (dashed line), and the magnetoplasmon
frequency is fitted either by Eq. (2) (solid curve) or by the
hydrodynamical model (dotted curve). Theoretical curves in (b) are
identical to that plotted in (a).} \label{Fig3}
\end{center}
\end{figure}

\begin{figure}
\begin{center}
\epsfig{file=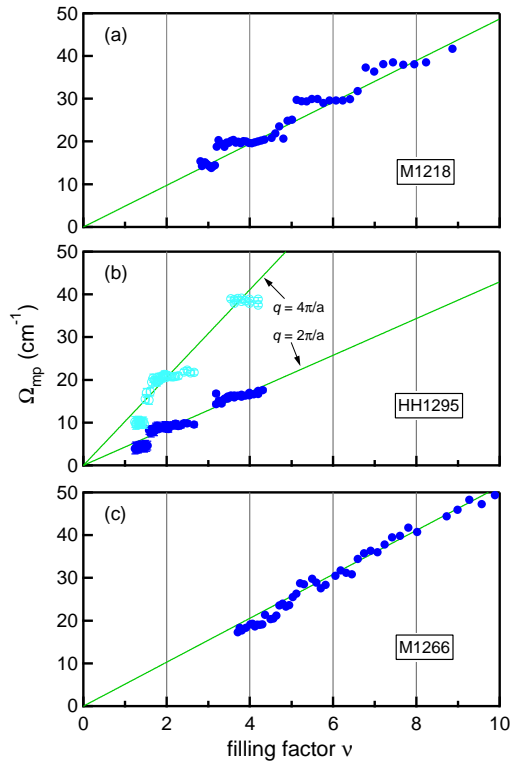,height=10.5cm} \caption{(color online).
Filling-factor dependence of the renormalized frequency
$\Omega_{mp}$ for the resistively-detected magnetoplasmons
measured in three different samples. The solid lines are the
semiclassical predictions calculated using Eq. (3), which fit
exactly the dispersions measured by the absorption experiments.}
\label{Fig4}
\end{center}
\end{figure}

In Fig. \ref{Fig3}(a) and (b) we plot the $B$-field dispersions of
the charge excitations determined from the absorption and FIR-PC
spectra, respectively. In both cases the CR can be well fitted
(dashed line) using the linear relation $\omega_{c} = eB/m^{\ast}$
with $m^{\ast}$ = 0.0726 $m_{e}$. Knowing the effective mass, we
fit (solid curve) in Fig. \ref{Fig3}(a) the dispersion of
magnetoplasmon measured by absorption spectroscopy using the eqs.
(1) and (2), and determine $q_{TF}$ = 1.83$\times 10^{6}$
cm$^{-1}$. Similar fitting procedures are performed for other
samples and the obtained values for $m^{\ast}$ and $q_{TF}$ are
summarized in the Table 1. Eqs. (1) and (2) capture well the
general feature of the magnetoplasmon dispersion except for the
nonlocal effect \cite{Batke}, which is responsible for the
anticrossing of the magnetoplasmon with the harmonics of CR (the
Bernstein modes) and an increase of the magnetoplasmon energy at
the low $B$ fields. Using the hydrodynamical model \cite{Chaplik}
taking into account the nonlocal effect and with the same
parameters of $m^{\ast}$ and $q_{TF}$, the calculated
magnetoplasmon dispersions (dotted curves) agree well with that
measured by absorption spectroscopy in the whole $B$-field range,
in accordance to previous studies \cite{Batke}. In contrast,
compared to the theoretical curves and the absorption experiment,
the magnetoplasmon dispersion measured by FIR-PC spectroscopy
shows obvious deviations in Fig. \ref{Fig3}(b). Plotted in this
scale that covers the entire CR frequency range, the deviation
looks small. In fact, it is well beyond the experimental accuracy.
However, before we demonstrate it more clearly in Fig. \ref{Fig4}
by plotting the renormalized magnetoplasmon frequency
$\Omega_{mp}$ in which the CR frequency is subtracted, let us
first check how important the problem is for deviation of the
magnetoplasmon energy by studying two questions: why does the
semiclassical magnetoplasmon dispersion describe the absorption
data so well? And what differs in the FIR-PC experiment?

The answer to the first question lies in the long wavelength of
the magnetoplasmons we investigate, for which $q\ell << 1$ at
large $B$ fields. Here $\ell = \sqrt{\hbar/eB}$ is the magnetic
length. Under this condition influences of quantum \cite{Chiu} and
correlation effects \cite{Kallin} on the magnetoplasmon are small.
Therefore, deviations of the magnetoplasmon dispersion from the
semiclassical prediction as shown in the Fig. \ref{Fig3}(b) are
unexpected and bring us to the second question.

In the absorption spectroscopy, one detects the elementary
excitations by measuring the transmitted radiation, assuming that
absorption of photons does not change the properties of the
electronic system. On the contrary in the PC experiments,
elementary excitations are detected by measuring the
photo-induced change of the resistance, which monitors exactly
the change of the electronic system caused by absorption of
photons. Only if the excited electronic system can reach a steady
state characterized by a slightly raised temperature, which is
known as the bolometric effect \cite{Neppl}, the same elementary
excitation will be detected in the PC experiments as by the
absorption spectroscopy. The bolometric model breaks down if
intense radiation drives the electronic system far beyond
equilibrium (as in the microwave PC experiments
\cite{Mani,Kukushkin,Zudov}), or if the energy relaxation is
either spatially inhomogeneous as in the QHE regime \cite{Kawano}
or is spin-dependent as for the spin-polarized electronic system
\cite{Zehnder}. All provide us chances for exploring unique
natures of elementary excitations unable to be investigated by
conventional absorption spectroscopy.

The relation between the QHE and the resistively detected
magnetoplasmon is clearly revealed in Fig. \ref{Fig4} in which we
summarize the filling-factor dependence of $\Omega_{mp}$
resistively measured on all our samples. For comparison,
semiclassical predictions for $\Omega_{mp}$ calculated by Eq. (3)
using the parameters of $m^{\ast}$ and $q_{TF}$ listed in Table 1
are plotted with solid lines. In Fig. \ref{Fig4}(a) $\Omega_{mp}$
measured on sample M1218 with the highest mobility deviates
clearly from the semiclassical prediction. Very interestingly, the
data shows plateaus forming around even filling factors of $\nu$ =
4, 6 and 8. In Fig. \ref{Fig4}(b) we plot $\Omega_{mp}$ obtained
on the sample HH1295 with a smaller density. The grating coupler
of this sample has a higher efficiency which allows us to measure
the magnetoplasmon modes at $q = 2\pi\cdot n/a$ with $n$ = 1
\textit{and} 2. Both show plateaus in the dispersion around the
even filling factors of $\nu$ = 2 and 4. The oscillatory behavior
is less obvious in the Fig. \ref{Fig4}(c) for the sample M1266
which has the lowest mobility.

The results shown in Fig. \ref{Fig4} are astonishingly reminiscent
of the celebrated QHE measured by DC magnetotransport
\cite{Chakraborty}, where the Hall conductivity equals to its
semiclassical prediction $\sigma_{H} = (e^{2}/h)\cdot \nu$ at even
filling factors (if the spin degeneracy is not lifted), with
plateaus forming around them. Decades after its discovery,
consensus for QHE has been established for the coexistence of the
compressible and incompressible strips with spatially varying and
constant electron density, respectively, which is the consequence
of the edge depletion and the strongly nonlinear screening
properties of the 2DES \cite{Chakraborty}. But confusion or
controversy remains existing regarding where the current flows in
a Hall bar \cite{Guven}. Dynamic properties are very helpful for
understanding the physics of QHE, however, investigations have so
far been focused on the edge magnetoplasmon mode that has the
one-dimensional character with energy much smaller than the
cyclotron energy \cite{Wassermeier}. Deviations of the
long-wavelength high-energy magnetoplasmon mode exhibiting
quantized features in the QHE regime have neither been predicted
nor been measured. Here we would follow the most recent
theoretical model of G\"{u}ven and Gerhardts which is improved by
Siddiki and Gerhardts (GGSG) \cite{Guven} to give a tentative
explanation of the quantized dispersion observed in Fig.
\ref{Fig4}.

On the basis of a quasi-local transport model including non-linear
screening effects on the conductivity, GGSG \cite{Guven} study the
current and charge distribution in the 2DES in the QHE regime.
Their model reproduces both longitudinal and Hall resistances with
exactly quantized Hall plateau. In contrast to previous pictures
\cite{Chakraborty}, GGSG find that a broad incompressible strip
exists at the center of the Hall bar at even filling factors. With
decreasing $B$, the incompressible strip moves from the center
towards the sample edges and disappears before the next is formed.
Assuming local equilibrium, they find that the total applied
dissipative current flows through the incompressible strips. This
theory provides an appealing basis for a simple explanation of the
$\Omega_{mp}$ plateau we observed: in contrast to the
semiclassical magnetoplasmon determined by the macroscopic motion
of all electrons in the sample that is measured by the absorption
spectroscopy, the resistively-detected magnetoplasmon is sensitive
to the local density of the incompressible strip. At exact even
filling factors its excitation energy is equal to the
semiclassical one. With decreasing $B$ field, the filling factor
in the incompressible strip is kept and therefore the
magnetoplasmon frequency deviates from the semiclassical
prediction and results in the plateau. GGSG also point out that
the effect of the broadening of the Landau levels reduces the
width of the incompressible strips, in accordance to the mobility
dependence of the plateau we observed. This simple interpretation
seems to match nicely the striking simplicity of the data shown in
Fig. \ref{Fig4}, and in principle PC spectroscopy is able to
detect local excitations in the QHE regime \cite{Merz}. However,
to explain the Hall plateau below the integer filling factors,
GGSG model requires the additional assumption of long-range
potential fluctuations, whose influence on the magnetoplasmon
below the integer filling factor needs to be examined. We note
that without a microscopic model, the question is open whether a
quasi-local plasmon modes may be stabilized in the incompressible
strips \cite{Mikhailov04}.

In summary, we have demonstrated directly by experiments that the
magnetoplasmon measured by photoconductivity and absorption
spectroscopy differs even under weak illumination. The dispersion
of the resistively-detected magnetoplasmon shows unexpected but
clearly-resolved plateaus for high-mobility 2DESs in the QHE
regime, which we interpret as effects of the dynamic response of
the incompressible strips. The unique effect provides a new basis
for developing a microscopic theory that could bring insight into
the dynamic properties of QHE as well as the nature of the
resistively-detected charge excitations in the photoconductivity
experiments, both are currently of great interest.

This work is supported by the EU 6th-Framework Programme through
project BMR-505282-1, the DFG through SFB 508 and BMBF through
project 01BM905. We thank R.R. Gerhardts, W. Hansen and S.A.
Mikhailov for helpful discussions and RRG for making
ref.\cite{Guven} available to us before publication.

\end{document}